\documentclass[preprint,showpacs,amsmath,amssymb,aps,prd]{revtex4}

\newcommand{\ket}[1]{{\mathop{\, |  #1  \rangle}}}

\newcommand{\del}{\partial}
\newcommand{\cA}{{\cal A}}
\newcommand{\cO}{{O}}

\newcommand{\prirodni}{\ensuremath{\mathbb{N}}}
\newcommand{\tM}{\tilde{M}}
\newcommand{\tN}{\tilde{N}}
\newcommand{\tS}{\tilde{S}}
\newcommand{\vn}{\vec{n}}

\newcommand{\dg}{{\rm dg}}
\newcommand{\ndg}{{\rm ndg}}
\newcommand{\Ddg}{{D_{\dg}}}
\newcommand{\Dndg}{{D_{\ndg}}}
\newcommand{\Rd}{{\mathop{{\,\rm Re\,}}\nolimits}}
\newcommand{\Imd}{{\mathop{{\,\rm Im\,}}\nolimits}}
\newcommand{\sgn}{{\mathop{{\,\rm sgn\,}}\nolimits}}

\begin{document}

\title{Graviton propagator asymptotics and the classical limit of  ELPR/FK spin
foam models}

\author{Aleksandar Mikovi\' c}
 \email{amikovic@ulusofona.pt}
\altaffiliation[Member of ]{Grupo de F\'isica Matem\'atica da Universidade de
Lisboa, Av. Prof. Gama Pinto, 2, 1649-003 Lisboa, Portugal}
\affiliation{Departamento de Matem\'atica, Universidade Lus\'{o}fona de
Humanidades e Tecnologia, Av. do Campo Grande, 376, 1749-024, Lisboa, Portugal}
\author{Marko Vojinovi\' c}
 \email{vmarko@cii.fc.ul.pt}
\affiliation{Grupo de F\'isica Matem\'atica da Universidade de Lisboa,
Av. Prof. Gama Pinto, 2, 1649-003 Lisboa, Portugal}

\date{\today}

\begin{abstract}
We study the classical limit of the ELPR/FK spin foam models by analyzing the
large-distance asymptotics of the corresponding graviton propagators. This is
done by examining the large-spin asymptotics of the Hartle-Hawking wavefunction
which is peaked around a classical flat spatial geometry. By using the
stationary phase method we determine the wavefunction asymptotics. The obtained
asymptotics does not give the desired large-distance asymptotics for the
corresponding graviton propagator. However, we show that the ELPR/FK vertex
amplitude can be redefined such that the corresponding Hartle-Hawking
wavefunction gives the desired asymptotics for the graviton propagator. 
\end{abstract}


\maketitle

\section{\label{SekI}Introduction}

Loop quantum gravity is a candidate for a realistic quantum theory of gravity
and it represents a nonperturbative and background independent way of quantizing
general relativity \cite{lqg}. However, one of its main problems is finding the
classical limit. This is difficult to do in the canonical formulation, because
there are no appropriate solutions of the Hamiltonian constraint. But even if
one had such a solution, it would be a complicated expression, and showing that
its transform to the triad representation has a semiclassical limit which
implies the Einstein equations is a daunting task, see \cite{Mikovic2004}. In
the covariant formulation,  i.e. the spin foam formalism, one can compute the
transition amplitudes between the spin network states, from which one can infer
the spin network wavefunction. However, this will be again a complicated
expression, and it will be difficult to compute the classical limit.

In spite of these difficulties Rovelli found a way to study the semiclassical
limit indirectly \cite{Rovelli2006}. His idea was to consider the graviton
propagator within the spin foam formalism and to study the semiclassical limit
by analyzing the large distance asymptotics of the propagator. By using an
assumption that the flat-space wavefunction has a specific Gaussian form in the
spin network basis, Rovelli was able to show that the graviton propagator had
the correct large distance asymptotics.  For more detailed studies and further
developments see \cite{Bianchi2006,Alesci2008,Mikovic2008,Bianchi2009}.

In \cite{Mikovic2008} it was pointed out that the Gaussian wavefunction which
had been used to calculate the graviton propagator asymptotics does not satisfy
the Hamiltonian constraint, also see \cite{Freidel}. Although the physical
wavefunction $\Psi_{\gamma}(j,j_0)$  is not a Gaussian, one will still obtain
the desired propagator asymptotics if $\Psi_{\gamma}(j,j_0)$  is approximated by
the Rovelli's Gaussian wavefunction for large spins, i.e.
\begin{widetext}
\begin{equation} \label{RovelliAnsatz}
\Psi_{\gamma}(j;j_0) \approx \cA(j_0) \exp \left[-\frac{1}{2j_0}\sum_{a,b}
\alpha_{ab}(j_a-j_0) 
(j_b -j_0) +
i\sum_a \theta_a  j_a\right]\,.
\end{equation}
\end{widetext}
Here $\gamma$ denotes the spin network graph, $j_a$ is a spin of a link of
$\gamma$, $\theta_a$ are arbitrary constants and $\alpha$ is a numerical matrix.
The parameter $j_0$ determines the scale of a triangle area in the spin network
and can be related to the boundary background metric, see \cite{Mikovic2008}.

However, nobody has investigated whether any viable candidate for the flat-space
wavefunction has the Gaussian asymptotic form (\ref{RovelliAnsatz}). Note that
such an analysis has been recently performed in the case of canonical Euclidean
loop quantum gravity (LQG) theory \cite{Mikovic2010}. It was shown that the
wavefunction has a Gaussian asymptotics, but it is not of the form
(\ref{RovelliAnsatz}). This result then implies that the Euclidean LQG graviton
propagator does not have the desired large-distance asymptotics. In order to be
sure about the implications of this result for physics, one needs to perform the
same analysis in the Lorentzian case. This can be done by using ELPR/FK spin
foam models \cite{elpr,fk}, since these are the only spin foam models that have
a Lorentzian formulation and give rise to a LQG theory on the spin foam
boundary.

In order to satisfy the Hamiltonian constraint, we will consider a boundary spin
network wavefunction obtained from the spin foam state sum for a spin foam with
a spin network boundary. This is a spin foam analog of the Hartle-Hawking
wavefunction \cite{HartleHawking}, and it is known that a Hartle-Hawking
wavefunction satisfies the Hamiltonian constraint. Guided by the construction of
the flat-space wavefunction in the Euclidean LQG case \cite{Mikovic2004}, we
will introduce the edge insertions in the boundary spin network in order to
simulate the presence of the boundary background metric. The large-spin
asymptotics of the boundary wavefunction will be studied  by using the
stationary phase method. We will determine the conditions necessary for the
asymptotics to be of the form (\ref{RovelliAnsatz}), see Eq.
(\ref{IntegraljenaTalasnaFunkcija}). Since the graviton propagator asymptotics
is determined by the $j_0$-dependence of the exponent in (\ref{RovelliAnsatz}),
we will focus our attention on the coefficient $S$ in
(\ref{IntegraljenaTalasnaFunkcija}), which is determined by the Hessian matrix
for the logarithm of the spin-foam amplitude for the boundary wavefunction. 

The method to determine the $j_0$-dependence of $S$ relies on a nontrivial
mathematical result formulated in Theorem 1. We obtain that $S=O(1)$, rather
than the desired result $S=O(1/j_0)$, see Eq. (\ref{ooa}). The result $S=O(1)$
implies that the graviton propagator behaves as the distance to the fourth power
in the limit of large distances. We will also show that the $S=O(1)$ asymptotics
is a direct consequence of the vertex amplitude asymptotics
(\ref{AsimptotikaVerteksa}), which is a common feature of all known spin foam
models.

This paper is organized as follows. In  section \ref{SekII} we introduce the
boundary spin-network wavefunction with insertions. In section \ref{SekIII} we
rewrite the wavefunction in the form suitable for the asymptotic analysis.
Section \ref{SekIV} is devoted to the analysis of the critical points of the
wavefunction, which play a major role in the asymptotic analysis. We discuss the
properties of the stationary point equations and outline a method that can be
used to solve them. However,  it is not necessary to solve explicitly the
stationary point equations since it is sufficient to use certain properties of
the critical points. In section \ref{SekV} we apply the extended stationary
phase method to determine the asymptotic behavior of the wavefunction in the
large-spin limit. A detailed analysis shows that if certain reasonable
assumptions are satisfied, the wavefunction will have a Gaussian asymptotics.
The width of the Gaussian is determined by a complex matrix which is essentially
the Schur complement of the Hessian of the logarithm of the integrand. It
depends in a nontrivial way on the scaling parameter $j_0$. In order to be able
to compare the wavefunction asymptotics with the Gaussian from
(\ref{RovelliAnsatz}), we need to determine the scaling of the Schur complement
in the limit $j_0\to\infty$, which is done in section \ref{SekVI}. An explicit
calculation of the Schur complement will not be possible, but it will be
possible to determine its scaling dependence on $j_0$. Surprisingly, one finds
that in the leading order the Schur complement scales as a constant in the limit
$j_0\to\infty$, in contrast to the assumed $1/j_0$ scaling in
(\ref{RovelliAnsatz}). This implies that the corresponding graviton propagator
does not have the distance scaling corresponding to a graviton propagator from
general relativity. In the final section \ref{SekZakljucak} we discuss the
possible ways to solve this problem and to recover the desired scaling of the
propagator. It turns out that the most promising method is to redefine the
vertex amplitude of the spin foam model, and we propose two ways to do that. The
appendices \ref{AppReggeDejstvo}, \ref{AppELPRFKasimptotika},
\ref{AppendixTeoremaSmatrice}, \ref{AppendixMatrixLemma} and
\ref{AppIzvodiVerteksa} contain derivations of the results that were used in the
main text.

\section{\label{SekII}The boundary wavefunction}

A boundary state $\ket{\Psi}$ for an ELPR/FK spin foam model can be constructed
in the following way. We expand $\ket{\Psi}$ in the spin network basis $
\ket{\gamma,j_l,\iota_p}$, where $\gamma$ is the boundary spin network graph,
$j_l$ are the spins of the edges of $\gamma$ and $\iota_p$ are the corresponding
intertwiners. We then expand each $ \ket{\gamma,j_l,\iota_p}$ in the coherent
state basis $\ket{\gamma,j_l,\vn_{pl}}$, see \cite{cohst}, so that
$$
\ket{\Psi} = \sum_{\gamma}\sum_{j_l} \int \prod_{(pl)} d^2\vn_{pl} \;
\Psi_{\gamma}(j_l,\vn_{pl};j_0)
\ket{\gamma,j_l,\vn_{pl}}\,.
$$
The coefficients $\Psi_{\gamma}(j,\vn;j_0)$ are constructed as boundary
spin-network wavefunctions with edge insertions. The edge insertions are
introduced in order to provide the wavefunction with the
information about the boundary background metric. This is done through the
background spin parameter $j_0$,  so that
\begin{widetext}
\begin{equation} \label{TalasnaFunkcija}
\Psi_{\gamma}(j_l,\vn_{pl};j_0) =  \prod_{l\in \gamma}
\mu_l(j_l;j_0) d_f(j_l) \sum_{\substack{ k_f \\ \del\sigma=\gamma }} \int
\prod_{(ef)} d^2\vn_{ef} \prod_f
d_f(k_f) \prod_v W_v (k_f, \vn_{ef},j_l,\vn_{pl})\, ,
\end{equation}
\end{widetext}
where  $\sigma$ is a 2-complex whose boundary one-complex is $\gamma$. The face
labels $k$ and the edge-face labels $\vn$ of the corresponding spin foam are
fixed to be $j_l,\vn_{pl}$ at the boundary spin network. $k_f$ is a non-boundary
spin, which labels a face $f$, while a unit vector $\vn_{ef}$ labels an edge $e$
and the face $(ef)$ adjacent to $e$ in the 2-complex $\sigma$. One can also
include a sum over various 2-complexes $\sigma$ that have the fixed boundary
$\gamma$ in (\ref{TalasnaFunkcija}), thereby implementing the ``sum over
triangulations'' idea. However, this will not affect our analysis, so that we
will work with a single $\sigma$ for a given $\gamma$. The expressions for the
face and the vertex amplitudes $d_f$ and $W_v$ can be found in
\cite{elpr,fk,Rovelli2010,Bianchi2010} and we do not write them explicitly
because we will need only their asymptotic form for large spins.

The motivation for the introduction of the edge insertions  $\mu_l(j_l ,j_0)$
comes from the construction of the flat-space wavefunction in the Euclidean
canonical LQG \cite{Mikovic2004}. This wavefunction solves the Hamiltonian
constraint and it is given by a state-sum similar to (\ref{TalasnaFunkcija}).
The parameter $j_0$ is proportional to the areas of the triangles determined by
the background geometry triads. The boundary spin network insertions are
arbitrary functions of the edge spins and $j_0$. Therefore the expression
(\ref{TalasnaFunkcija}) is a natural  generalization of the Euclidean
wavefunction from \cite{Mikovic2004} to the Lorentzian geometry case.
Furthermore, since (\ref{TalasnaFunkcija}) is constructed as a Hartle-Hawking
wavefunction for a boundary spin network for the ELPR/FK spin foam model, it
will satisfy automatically the corresponding Hamiltonian constraint and the
insertions will insure that it is peaked around a flat spatial geometry. Since
the insertions $\mu_l(j_l ,j_0)$ can be arbitrary functions, one can try to
choose them such that the asymptotics (\ref{RovelliAnsatz}) is obtained.

The vectors $\vn_{ef}$ are in general defined up to arbitrary phase factors.
These phase factors can be chosen such that they insure nice gluing properties
of neighboring simplices in the triangulation dual to $\sigma$. As discussed in
\cite{Barret2009eeprl}, such a choice will fix the phase factors on the spin
foam boundary $\gamma$, and thus give rise to the phase term in
(\ref{RovelliAnsatz}). However, these phase factors disappear when the graviton
propagator is calculated in the standard canonical formalism, see
\cite{Mikovic2008}, and therefore their values will not be important for our
purposes.

\section{\label{SekIII}Asymptotic analysis}

The wavefunction (\ref{TalasnaFunkcija}) does not necessarilly have the
large-spin asymptotic form (\ref{RovelliAnsatz}). In what follows, we are going
to study its large-spin asymptotics, in order to find out is there a choice of
the insertions such that the asymptotics (\ref{RovelliAnsatz}) is obtained. If
the asymptotics is indeed of the form (\ref{RovelliAnsatz}), the wavefunction
(\ref{TalasnaFunkcija}) can be a good candidate for a flat-space wavefunction.

We begin the analysis of the large-spin asymptotics of (\ref{TalasnaFunkcija})
by defining the large-spin limit. Namely, we are interested in the limit
\begin{equation} \label{SkaliranjeSpinova}
j_l = j_0 \tilde{\jmath}_l, \qquad j_0\to \infty.
\end{equation}
Here $\tilde{\jmath}_l\in\prirodni_0/2$ are spins which are fixed, while $j_0$
is the large parameter.

It is important to note that the scaling of boundary spins $j_l$ via the
parameter $j_0$ will induce a similar scaling in some of the internal spins
$k_f$, due to the triangle inequalities built in the vertex amplitude $W_v$.
However, not all internal spins need to be scaled, depending on the
combinatorics of the two-complex $\sigma$. The domain of summation in
(\ref{TalasnaFunkcija}) will contain sectors where all spins are scaled and
sectors where only some of them are scaled. Those internal spins which must
scale do so by a prescription analogous to (\ref{SkaliranjeSpinova}).

The first step in finding the asymptotic behavior of (\ref{TalasnaFunkcija}) is
to approximate the sums over the internal spins $k_f$ with integrals. The
wavefunction (\ref{TalasnaFunkcija}) can  be then approximated as
$$
\Psi_{\gamma}(j_l,\vn_{pl};j_0) \approx  I_\gamma (j_l,\vn_{pl};j_0) =
\hphantom{mmmmmmmmm}
$$
\begin{equation} \label{EksponencijalniOblikTF}
\hphantom{mmmmm}
= \int_D
\prod_{f} dk_f \int \prod_{(ef)}
d^2\vn_{ef} \, e^{j_0 F(j,k,\vn;j_0)}\,,
\end{equation}
where the function $F$ is given by
$$
F(j,k,\vn;j_0) = \frac{1}{j_0} \sum_l \log \mu_l(j_l;j_0) d_l(j_l) +
$$
\begin{equation} \label{Dejstvo}
+\frac{1}{j_0} \sum_{f\neq l} \log d_f(k_f) +
\frac{1}{j_0} \sum_v \log W_v(j,k,\vn)
\, .
\end{equation}
$D$ is the domain of integration over spins $k$ and the form
(\ref{EksponencijalniOblikTF}) is suitable for the stationary phase
approximation. Note that the vertex amplitude $W_v$ is complex-valued in
general, so that the logarithm is defined up to a multiple of $2\pi i$. However,
this constant factor does not influence the subsequent analysis and we can
ignore it. Also note that the insertion functions $\mu_l$ depend explicitly on
$j_0$, while $d_f$ and $W_v$ may depend on $j_0$ only through boundary spins $j$
and those internal spins $k$ that are constrained to scale via triangle
inequalities.

We will use the extended stationary phase method \cite{Hormander1983} in order
to approximate the integral (\ref{EksponencijalniOblikTF}). The method will be
applicable if the function (\ref{Dejstvo}) satisfies
\begin{equation} \label{UslovZaMetodStacionarneTacke}
F(j,k,\vn;j_0) = \cO(1),
\end{equation}
for $j_0\to\infty$. This condition will be satisfied on a subset of $D$ where
the asymptotic formulae for the ELPR/FK vertex amplitude $W_v$, derived in
\cite{Barret2009ooguri,Barret2009eeprl,Barret2009leprl}, are valid.  See  the
appendix \ref{AppELPRFKasimptotika} for the explicit expressions.

When the boundary spins $j_l$ are large, i.e. $j_l = O(j_0)$, then the
integration domain $D$ will contain spin foams whose spins are all large. $D$
will also contain spin foams where some of the spins are large and other are
small. This structure is a consequence of the triangular inequalities  among the
spins which form a spin-foam vertex (rules for the addition of angular momenta).
Let $\Dndg$ be the set of spin foams in $D$ such that each spin foam from
$\Dndg$ contains at least one vertex with all spins large. Then  $\Ddg = D
\setminus \Dndg$ is the set of spin foams where every vertex in a spin foam from
$\Ddg$ contains a small spin. Consequently
$$
I_{\gamma} = I_{\gamma}^{\ndg}+ I_{\gamma}^{\dg}\, ,
$$
where $I_{\gamma}^{\ndg}$ and $I_{\gamma}^{\dg}$ are defined by taking the
integral (\ref{EksponencijalniOblikTF}) over the  domains $\Dndg$ and $\Ddg$,
respectively.

It is not known whether the function $F$ satisfies the condition
(\ref{UslovZaMetodStacionarneTacke}) on $\Ddg$, since the asymptotic formulae
for $W_v$ when some of the vertex spins are large and the other are small are
not known. On the other hand, the asymptotic formula for $W_v$ in the case when
all the vertex spins are large is known, see  (\ref{AsimptotikaVerteksa}) and
(\ref{jav}), so that it can be shown that $F$ satisfies the condition 
(\ref{UslovZaMetodStacionarneTacke}) on $\Dndg$. This is true because every spin
foam from $\Dndg$ contains at least one vertex with nondegenerate asymptotics,
and  therefore the  contribution of such a vertex to $F$ is given by
\begin{widetext}
$$
\frac{1}{j_0} \log W_v^{\ndg} \approx  \frac{1}{j_0}
\log\left( N_+^{(\alpha)}e^{i\alpha S_R^{(v)}} +N_-^{(\alpha)}
e^{-i\alpha S_R^{(v)}} \right) +
O\left(\frac{ \ln j_0}{ j_0}\right)  \, .
$$
Since $N_\pm \ne 0$, then
\begin{equation}\frac{1}{j_0} \log W_v^{\ndg} \approx  i\alpha
\frac{S_R^{(v)}}{j_0} + \frac{1}{j_0}
\log\left( N_+^{(\alpha)} +N_-^{(\alpha)}
e^{-2i\alpha S_R^{(v)}} \right) +O\left(\frac{ \ln j_0}{  j_0}\right)  \,
.\label{ndgvc}
\end{equation}
\end{widetext}
According to (\ref{SkaliranjeReggeDejstva}) the Regge action $S_R^{(v)}$ is of
$O(j_0)$, so  that the first term in (\ref{ndgvc})  is of $O(1)$. Since the
coefficients $N_{\pm}^{(\alpha)}$ are of $O(1)$, the second term  in
(\ref{ndgvc})  is of $O(j_0^{-1})$ . Therefore, a nondegenerate vertex gives an
$O(1)$ contribution to the function (\ref{Dejstvo}). 

A degenerate vertex from $\Dndg$  can give a  contribution to $F$ of $O(1)$ or
lower, depending on the type of degeneracy of each particular vertex. The sum
over the insertion functions $\mu_l$ in $F$ can be chosen such that it is of
$O(1)$. One particularly useful choice for the insertion functions is
\begin{equation} \label{MuFunkcije}
\mu_l(j_l;j_0) = \exp\left[-\frac{(j_l-j_0)^2}{j_0}\right] \, .
\end{equation}
This choice is very natural for our purposes, since it enforces the flat
background metric in the boundary state and it gives an $O(1)$ contribution to
$F$. As far as the the  sum over the face amplitudes in $F$ is concerned, it is
of  $O(j_0^{-1}\ln j_0)$, which is subleading to $O(1)$. This is because $d_f
(j)$ is of $O(j_0^q)$, where $q=1$ or $q=2$, see \cite{Bianchi2010} for a
discussion of the various proposals for $d_f (j)$. Therefore $F=O(1)$ on 
$\Dndg$,  provided that there is no cancellation of $O(1)$ terms. Hence one can
use the stationary phase approximation for the integral $I^{\ndg}$.

As far as the order of $F$ on $\Ddg$ is concerned, it can be of $O(1)$ if the
choice (\ref{MuFunkcije}) is used. However, the stationary phase approximation
cannot be made because of the absence of the asymptotic formulas for the
degenerate vertices.

Also note that the extended stationary phase method is directly applicable only
if $F$ is a Morse function, which means that its Hessian matrix does not have
zero eigenvalues at the critical points. However, in our case the Hessian of $F$
may happen to be degenerate, so that we need to take this fact into account when
applying the stationary phase method. This will be discussed in detail in
section \ref{SekV}.

\section{\label{SekIV}Critical points}

The idea of the stationary phase method is to approximate the integrand in the
nondegenerate piece of $I_\gamma$ as a sum of Gaussian functions, where each
Gaussian is centered around a stationary point $(j^*,k^*,\vn^*)$ of $e^{j_0 F}$.
As $j_0\to\infty$, only the immediate neighborhoods of the stationary points
will contribute to the integral \cite{Hormander1983}. Furthermore, only the
stationary points for which 
\begin{equation}
\Rd F (j^*,k^*,\vn^*;j_0) = 0\,,\label{crc} 
\end{equation}
will give a noticeable contribution. The stationary points which satisfy
(\ref{crc}) are called the critical points.

Note that the stationary points of $e^{j_0 F}$ are the same as the stationary
points of $F$. Therefore, the stationary point equations are given by
\begin{equation}
\frac{\del F}{\del j_l}=0\,, \qquad \frac{\del F}{\del k_f} =0\,, \qquad
\frac{\del F}{\del \vn_{ef}}=0\,.\label{stpe}
\end{equation}

The geometric interpretation of these equations has been studied extensively in
\cite{ConradyFreidel} for the Euclidean theory and in \cite{MagliaroPerini} for
both Euclidean and Lorentzian versions of the theory. In this paper we do not
need to go into the details, just let us mention that the condition (\ref{crc})
and the the stationary point equation for $\vec{n}_{ef}$ are satisfied only if
one considers a spin foam which is dual to the triangulation of a Regge
geometry. As far as the $j$ and $k$ equations are concerned, we can unify them
by using a common label $x_a$ for $j_l$ and $k_f$. Then $\frac{\del F}{ \del
x_a} =0$ gives
$$
\sum_{v\in\ndg } \frac{ N_+^{(\alpha)} e^{i\alpha S_R^{(v)}} - N_-^{(\alpha)}
e^{-i\alpha S_R^{(v)}} }{ N_+^{(\alpha)} e^{i\alpha S_R^{(v)}} + N_-^{(\alpha)}
e^{-i\alpha S_R^{(v)}}}
i\alpha \delta_{a\in v} \Theta_a^{(v)} + 
$$
$$
+ \sum_{v\in\ndg } \frac{ e^{i\alpha S_R^{(v)}} \del_a N_+^{(\alpha)} +
e^{-i\alpha S_R^{(v)}} \del_a N_-^{(\alpha)} }{ N_+^{(\alpha)}
e^{i\alpha S_R^{(v)}} + N_-^{(\alpha)} e^{-i\alpha S_R^{(v)}}} + 
$$
\begin{equation} \label{StacionarnaJednacinaZaJiK}
+ \sum_{v\in\dg} \del_a \log W_v
+ \sum_{l} \del_a \log\mu_l
+ \sum_f \del_a \log d_f = 0 \, ,
\end{equation}
where we have used  (\ref{IzvodReggeDejstva}). When $x_a = k_f$ then the $\mu_l$
terms are absent in (\ref{StacionarnaJednacinaZaJiK}), while for $x_a = j_l$ the
last sum in (\ref{StacionarnaJednacinaZaJiK}) goes only over the boundary $f$'s.

The $j_0$-dependence of (\ref{StacionarnaJednacinaZaJiK}) is the following. The
first sum in (\ref{StacionarnaJednacinaZaJiK})  is of  $O(1)$, and it gives the
dominant contribution to the equation. The second sum in
(\ref{StacionarnaJednacinaZaJiK}) is of  $O(1/j_0)$ since the factors
$N_{\pm}^{(\alpha)}$ are of  $O(1)$ and consequently their derivatives with
respect to $j$ and $k$ are of $O(1/j_0)$. The fourth and the fifth sum in
(\ref{StacionarnaJednacinaZaJiK}) are of $O(1/j_0)$, while the third sum is at
most of $O(1)$. Given this, the critical point equations (\ref{crc}) and
(\ref{stpe}) can be solved by restricting to a Regge triangulation and writing a
stationary point $x^*$ as
\begin{equation}
x^*_a = c_a j_0 + d_a + O(1/j_0) \,,\label{stpa}
\end{equation}
where $c_a$ and $d_a$ are coefficients to be determined. The equations can be
then expanded into a power series in $1/j_0$ and solved order by order for $c$
and $d$. This has been done explicitly in \cite{Mikovic2010} for the case of
Euclidean LQG flat-space wavefunction. In that case the vertex amplitude is the
$6j$ symbol, and some explicit solutions can be found.

However, for the purposes of this paper, it is not necessary to construct
explicit solutions of (\ref{crc}) and (\ref{stpe}). Rather, we only need to
assume that they have at least one nontrivial solution $(j^*,k^*,\vn^*)$ which
is a critical point. If there are no such solutions, the integral
$I^{\ndg}_{\gamma}$  will be of $o(1/j_0^n)$ for all $n>0$, and thus it will not
have the asymptotic form (\ref{RovelliAnsatz}).

Note that  $\Theta_f^{(v)}= 0$ for all $f$  is a leading-order solution of
(\ref{StacionarnaJednacinaZaJiK}),  since if we neglect $O(1/j_0)$ terms we
obtain
$$
\sum_{v\in\ndg } \frac{ N_+^{(\alpha)} e^{i\alpha S_R^{(v)}} - N_-^{(\alpha)}
e^{-i\alpha S_R^{(v)}} }{ N_+^{(\alpha)} e^{i\alpha S_R^{(v)}} + N_-^{(\alpha)}
e^{-i\alpha S_R^{(v)}}}
i\alpha \delta_{a\in v} \Theta_a^{(v)} = 0 \, .
$$
However, such solutions have to be discarded because the corresponding
$(j^*,k^*,\vn^*)$ do not satisfy the triangular inequalities. The reason is that
the angles $\Theta_f^{(v)}$ are exterior dihedral angles of a 4-simplex dual to
$v$. Since a 4-simplex is a convex body, its exterior dihedral angles cannot all
be equal to zero.

\section{\label{SekV}Extended stationary phase method}

We are going to determine the large-spin asymptotics of $I^{\ndg}_\gamma$ by
using the extended stationary phase method. As explained in the previous
section,  we will assume  that there is a dominant critical point and such a
point must satisfy
\begin{equation}
\Theta_f^{(v)}\neq 0 \,,\label{tnez}
\end{equation}
for some $f$ and all $v$.

As we have already pointed out, the function $F$ may not be a Morse function,
and consequently we cannot apply directly the well known results
\cite{Hormander1983} so we will  perform the calculation step by step.

Let  $J =(j_l ,\vn_{pl})$, $K = (k_f ,\vn_{ef})$  and  $ x= (j_l$, $k_f$,
$\vn_{pl}$, $\vn_{ef})$. We first  approximate the integrand $e^{j_0 F}$ with a
sum of Gaussian functions, each centered around a critical point $x^*$. The
corresponding exponents are obtained by expanding $F$ into a power series around
each $x^*$ up to quadratic terms. The integral $I_\gamma^{\ndg}$ then becomes
$$
I_{\gamma}^{\ndg}(J;j_0) \approx \sum_{x*} e^{j_0 F^*} \int dK\, e^{\frac{1}{2}
(x-x^*)^T
\Delta(x^*) (x-x^*) } =
$$
\begin{equation} \label{PsiNedegPreIntegracije}
 = \sum_{x^*} e^{j_0 F^*} I^*(J,x^* ) \,.
\end{equation}
Here $F^* = F(x^*) = i\Imd F(x^*)$ is the evaluation of $F$ at the critical
point $x^*$, the sum goes over the set of all distinct critical points, and
$\Delta$ is the Hessian matrix of $j_0 F$ evaluated at $x^*$
\begin{equation} \label{DefinicijaDeltaMatrice}
 \Delta_{ab} \equiv j_0 \frac{\del^2 F}{\del x_a \del x_b} \Big|_{x^*}\,.
\end{equation}
The scaling parameter $j_0$ has been absorbed into $\Delta$ for convenience.

In order to perform the  Gaussian integrations in
(\ref{PsiNedegPreIntegracije}), we will split the $\Delta$ matrix into $JJ$,
$JK$ and $KK$ blocks, which will be denoted as $A$, $N$ and $M$, respectively
\begin{equation} \label{BlokoviDeltaMatrice}
\Delta = \left[
\begin{array}{cc}
A & N \\
N^T & M \\
\end{array}
 \right] .
\end{equation}

Let us rewrite the exponent in a Gaussian integral as
$$
\frac{1}{2} (x-x^*)^T \Delta (x-x^*) = \frac{1}{2} (J-J^*)^T A (J-J^*) +
$$
$$
+ \frac{1}{2} (K-K^*)^T M (K-K^*) + (J-J^*)^T N (K-K^*) \, .
$$
The first term is independent of $K$ and can be moved in front of the integral.
By making a change of variables $K= Q\hat{K}$ and by making a suitable choice of
the matrix $Q$, the matrix $Q^TMQ$  becomes diagonal. Then
$$
I^* =\prod_a I^*_a =  \prod_a \int_{D_a} d\hat{K}_a \, e^{\frac{1}{2} m_a
\hat{K}_a^2 + n_a \hat{K}_a}
\, .
$$
where $m_a$ are the eigenvalues of the matrix $M$, $n_a = \left[(J-J^*)^T N Q
\right]_a $, and $D_a$ is the one-dimensional domain of integration, determined
by $D$ and the change of variables $K=Q\hat K$.

Let us now discuss the integral $I_a^*$.
\begin{itemize}
\item If $m_a \neq 0$ and $\Rd m_a \leq 0$, the integral converges. Since
$m_a=O(j_0)$, in the limit $j_0\to \infty$ the result is independent of the
domain $D_a$ in the leading order, and can be written as
$$
I_a = e^{-\frac{n_a^2}{2m_a}} \sqrt{\frac{2\pi}{-m_a}} \left[ 1+
\cO\left(\frac{1}{j_0}\right) \right] \, .
$$
Note that quadratic dependence on $n_a$ generates a term of type $(J-J^*)^2$ in
the exponent, which gives us the desired Gaussian asymptotics.

\item If $m_a \neq 0$ and $\Rd m_a > 0$, the integral diverges exponentially in
the limit $j_0\to\infty$, so that in this case the Gaussian asymptotics cannot
be obtained.

\item If $m_a = 0$ and $n_a \neq 0$, the integral might or might not converge,
depending on whether the domain $D_a$ is compact or not. However, even when it
converges, the result will be a non-Gaussian function of $J-J^*$.

\item If $m_a = 0$ and $n_a = 0$, the integral converges if $D_a$ is compact.
Most importantly, in this case the result is independent of $J-J^*$. Namely, if
we denote $D_a = [\alpha_a,\beta_a]$, we have
$$
I_a = \beta_a - \alpha_a \equiv \cA_a \, .
$$
The integrals of this type do not influence the propagator asymptotics.
\end{itemize}

Therefore the integral $I^*$ will be a Gaussian function of $J-J^*$, if the
following conditions are satisfied:
\begin{itemize}
\item all nonzero eigenvalues of $M$ have their real part negative or zero,
\item the matrix $N$ is projected to zero on the kernel of $M$, and
\item the domain of integration over the kernel space of $M$ is compact.
\end{itemize}

These conditions imply that we can always make a change of variables such that
the matrices $M$ and $N$ are given by
$$
M = \left[
\begin{array}{cc}
\tM & 0 \\
  0 & 0 \\
\end{array}
 \right] \, , \qquad
N = \left[
\begin{array}{cc}
\tN & 0 \\
\end{array}
 \right] \, ,
$$
where the matrix $\tM$ is invertible and has a negative-definite real part, and
the horizontal dimension of the zero-block in $N$ is equal to the corresponding
dimension of the zero-block in $M$.

We note here that these zero blocks appear because the function $F$ has
continuous symmetries, which give rise to manifolds of stationary points instead
of discrete sets of stationary points. As we have shown, integrating over these
manifolds does not affect the propagator asymptotics, as long as they are
compact (otherwise the integral will diverge).  A similar situation was
encountered when applying the extended stationary phase method to the case of an
Euclidean spin foam model \cite{ConradyFreidel}, as well as to the case of a
spin foam consisting of a single vertex
\cite{Barret2009eeprl,Barret2009leprl,Barret2009ooguri}.

Consequently we obtain
\begin{equation} 
I^{\ndg}_{\gamma}(J;j_0) \approx \sum_{x^*} \cA(x^*,j_0) \, e^{\frac{1}{2}
(J-J^*)^T
\tS(x^*,j_0)
(J-J^*) } \,,\label{gsum}
\end{equation}
where 
$$
\cA(x^*,j_0) = e^{j_0 F^*} \sqrt{\frac{(-2\pi)^r}{\det \tM}} \prod_a \cA_a
\left[ 1
+ O\left( \frac{1}{j_0} \right) \right] \, .
$$
$r$ is the dimension of $\tM$, while the product is taken over $a$ for which
$m_a = 0$. Also,
\begin{equation} \label{SchurovKomplement}
\tS = A - \tN \tM{}^{-1} \tN^T
\end{equation}
is the Schur complement \cite{Zhang2005} of the (regular minor of the) Hessian
matrix $\Delta$.

Note that the asymptotic form (\ref{gsum}) is a sum of many Gaussian functions,
while the desired  asymptotics  (\ref{RovelliAnsatz}) is just a single Gaussian
function. The expression (\ref{gsum}) can yield a single Gaussian if there is a
dominant critical point $x_0^*$ such that  $|\cA(x_0^*)|$ dominates any other
$|\cA(x^*)|$  when $j_0\to \infty$. This point can be determined as the one for
which the dimension $r$ of $\tM$ is minimal, i.e. when $M$ is maximally
degenerate. Consequently 
\begin{equation} \label{IntegraljenaTalasnaFunkcijaStara}
I_{\gamma}^{\ndg}(J;j_0) \approx \cA (j_0) \, e^{\frac{1}{2} (J-J^*)^T \tS(j_0)
(J-J^*) } \,
\end{equation}
when $j_0 \to\infty$. Finally, the quadratic form in the exponent of
(\ref{IntegraljenaTalasnaFunkcijaStara}) can be decomposed into a sum of $jj$,
$j\vn$ and $\vn\vn$ terms. Since $j=\cO(j_0)$ and $\vn = \cO(1)$ then the $jj$
terms will be dominant in the limit $j_0\to\infty$. Therefore
\begin{equation} \label{IntegraljenaTalasnaFunkcija}
I_{\gamma}^{\ndg}(j;j_0) \approx \cA (j_0) \, e^{\frac{1}{2} (j-j^*)^T S(j_0)
(j-j^*) } \,,
\end{equation}
where $S$ is the $jj$ block of the Schur matrix $\tS$. Note that (\ref{stpa})
implies $j^* = O(j_0)$, so that (\ref{IntegraljenaTalasnaFunkcija}) can be
written as
\begin{widetext}
\begin{equation} \label{andgtf}
I_{\gamma}^{\ndg}(j;j_0) \approx \cA (j_0) \exp\left[\frac{1}{2}
\sum_{a,b}S_{ab}(j_0)(j_a- c_a j_0)  
(j_b - c_b j_0) \right] \,.
\end{equation}
\end{widetext}

From (\ref{IntegraljenaTalasnaFunkcija}) it follows that
\begin{equation}
\Psi_\gamma \approx  \cA (j_0) \, e^{\frac{1}{2} (j-j^*)^T 
S(j_0)
(j-j^*) } + I_\gamma^{\dg} \,.\label{nsga}
\end{equation}
In order to obtain a single Gaussian asymptotics we need to assume that
$I^{\dg}_{\gamma}$ has a subleading asymptotics to that of $I^{\ndg}_{\gamma}$.
Since we do not know how to calculate the asymptotics of $I^{\dg}_{\gamma}$,
there is a possibility that $I^{\dg}_{\gamma}$ has the right asymptotics which
is dominant with respect to $I^{\ndg}_{\gamma}$. However, we will argue that
such a possibility would give a classical limit whose spacetime geometry has the
curvature which varies greatly on small scales, while the corresponding
propagator is that for a flat spacetime, see section \ref{SekZakljucak}.
Therefore we will assume that (\ref{nsga}) implies
\begin{equation}
\Psi_\gamma \approx  \cA (j_0) \, e^{\frac{1}{2} (j-j^*)^T 
S(j_0)
(j-j^*) } \,.\label{sga}
\end{equation}

However, the asymptotics (\ref{sga}) is still not the desired asymptotics
(\ref{TalasnaFunkcija}). We need to to determine whether or not $S =\cO(1/j_0)$.
We will analyze this problem in the next section.

\section{\label{SekVI}Calculation of the exponent factor}

The asymptotic form (\ref{sga}) will give the desired asymptotics if
$S=O(1/j_0)$. Note that it is very difficult to calculate the matrix $S$
explicitly. However, we only need to calculate the leading $j_0$-order of $S$.
This can be done by using the following theorem

\medskip

\textbf{Theorem 1.} \textit{If the matrix $S$ is nonzero, and if the leading
order contribution to $\Delta$ comes from $W_v$ terms, we have}
$$
S = O(\Delta)
$$
\textit{for $j_0 \to\infty$. If the matrix $S$ is zero, the wavefunction
asymptotics is non-Gaussian.}

\medskip

The proof is essentially based on the Schur determinant formula, $\det \Delta =
\det S \det M$, see \cite{Zhang2005}, and is given in Appendix
\ref{AppendixTeoremaSmatrice} (see also \cite{Mikovic2010}).

The asymptotic dependence of $\Delta$ on $j_0$ can be determined quite easily,
if the large spin asymptotics of the vertex amplitude $W_v$ is known. From
(\ref{Dejstvo}) and (\ref{DefinicijaDeltaMatrice}) it follows that
\begin{equation} \label{OpstiOblikDeltaMatrice}
\Delta_{ab} = \sum_V \frac{\del^2 \log A_V}{\del x_a \del x_b} \,,
\end{equation}
where $V\in \{ l,f,v \}$, $ A_l = \mu_l$, $ A_f = d_f$, $ A_v = W_v,$ and the
derivatives are evaluated at the critical point $x^*$.  Each term in
(\ref{OpstiOblikDeltaMatrice}) contributes to the asymptotics of $\Delta$ with
some power of $j_0$, so that the leading order asymptotics of $\Delta$ will be
determined by the highest power of $j_0$.

The insertion functions can be chosen arbitrarily and therefore can give any
desired contribution of $O(j_0^p)$. However, they only contribute to the
diagonal elements of $\Delta$, since each insertion function $\mu_l$ depends
only on the spin of its link, $j_l$. For the choice (\ref{MuFunkcije}) one
easily gets from (\ref{OpstiOblikDeltaMatrice})
$$
\frac{\del^2 \log \mu_l}{\del x_a \del x_b}  =
-\frac{2\delta_{ab}\delta_{al}}{j_0} = O\left( \frac{1}{j_0} \right) \,.
$$

The face amplitude $d_f$ is commonly chosen to be $d_f(j)= 2 j_f + 1$, see
\cite{Bianchi2010}. Substituting into (\ref{OpstiOblikDeltaMatrice}), we obtain
$$
\frac{\del^2 \log d_f}{\del x_a \del x_b}  =
-\frac{4\delta_{ab}\delta_{af}}{(2x_f+1)^2} = O\left( \frac{1}{j_0^2} \right),
$$
and this is also a contribution to the diagonal elements of $\Delta$. Note that
other choices for $d_f(j)$ have also been proposed in the literature, see for
example \cite{fk}. However, all the proposed choices satisfy $d_f = O(j_f^q)$,
where $q\ge 1$, so that one obtains an $O(j_0^{-2})$ contribution.

Finally, the main nontrivial contribution comes from the vertex amplitude $W_v$.
The asymptotics of the degenerate configurations of the vertex amplitude is
unknown, and such vertices can in general give a contribution to $\Delta$ of
order $\cO(1)$ or smaller. However, the asymptotics of the nondegenerate
vertices is well studied, see Appendix \ref{AppELPRFKasimptotika}. Furthermore,
each spin foam in $\Dndg$ contains at least one nondegenerate vertex. By a
straightforward calculation one obtains from (\ref{OpstiOblikDeltaMatrice}) and
(\ref{jav})
\begin{equation} \label{DeltaV}
\frac{\del^2 \log W_v}{\del x_a \del x_b}  = \Delta_{vab}^{(0)} +
\Delta_{vab}^{(1)}
+ \Delta_{vab}^{(2)} \, ,
\end{equation}
where the three terms on the right-hand side represent contributions of order
$\cO(1)$, $\cO(j_0^{-1})$ and $\cO(j_0^{-2})$. In the non-Regge cases the
contributions are of $O(1/j_0^2)$ or subleading. The leading term is
\begin{widetext}
\begin{equation} \label{DeltaNula}
\Delta_{vab}^{(0)} =
 \left[ \left(
\frac{N_+^{(\alpha)} e^{i\alpha S_R^{(v)}} - N_-^{(\alpha)} e^{-i\alpha
S_R^{(v)}}}
{N_+^{(\alpha)} e^{i\alpha S_R^{(v)}} + N_-^{(\alpha)} e^{-i\alpha S_R^{(v)}}}
\right)^2 -1 \right] \alpha^2 \delta_{a,b\in v} \Theta_a^{(v)} \Theta_b^{(v)} \,
,
\end{equation}
\end{widetext}
while explicit expressions for $\Delta_{vab}^{(1)}$ and $\Delta_{vab}^{(2)}$ are
given in Appendix \ref{AppIzvodiVerteksa}.

The equation (\ref{DeltaV}) is evaluated at some critical point $x^*$. As we
have discussed in Section \ref{SekIV}, there are no critical points where all
the angles $\Theta_a^{(v)}$ are zero. Moreover, for the ELPR/FK vertex amplitude
the coefficients $N_+^{(\alpha)}$ and $N_-^{(\alpha)}$ are also always different
from zero, so the square bracket in (\ref{DeltaNula}) is also nonzero. Therefore
we see that the term (\ref{DeltaNula}) is always nonzero. Hence the dominant
contribution to the Hessian $\Delta$ comes from the vertex amplitudes whose
spins form a Regge geometry, and it is of order $\cO(1)$. Consequently, the
assumptions of Theorem 1 are satisfied, so that
\begin{equation} \label{ooa}
S = O(1) \, .
\end{equation}

The implication  of  (\ref{ooa}) for the large-distance asymptotics of the
graviton propagator can be seen from the following result. Consider a
generalized Rovelli asymptotics for the wavefunction
\begin{widetext}
\begin{equation} \label{RovelliAnsatzGeneralized}
\Psi_{\gamma}(j;j_0) \approx \cA(j_0) \exp \left[-\frac{1}{2j_0^p}
\sum_{a,b}\alpha_{ab}(j_a - c_a j_0)
(j_b - c_b j_0) \right]\,,
\end{equation}
\end{widetext}
where $p\geq 0$. Note that the obtained result (\ref{ooa}) corresponds to $p=0$
while the Rovelli ansatz corresponds to $p=1$. The propagator asymptotic scaling
with spacetime distance $|x-y|$ can be determined by repeating the calculation
done in \cite{Rovelli2006}, also see \cite{Mikovic2008}. Therefore one obtains
for large distances
\begin{equation}
G (x,y) \approx \frac{const}{|x-y|^{4-2p}} \,,\label{gpa}
\end{equation}
where $G$ denotes the diagonal components of the graviton propagator.

The equation (\ref{gpa}) gives for $p=1$ the propagator asymptotics consistent
with general relativity, while for $p=0$  it  gives 
\begin{equation}
G (x,y) \approx \frac{const}{|x-y|^{4}} \,.\label{elprpa}
\end{equation}
The asymptotics (\ref{elprpa}) is not consistent with general relativity.

\section{\label{SekZakljucak}Discussion and conclusions}

The result (\ref{ooa}) has been derived under certain assumptions, so that one
would like to know is it possible to relax the assumptions such that the desired
classical limit is obtained. The first thing one can try is to change the
insertion functions $\mu_l$, since these functions can be chosen freely. The
insertion functions could be chosen such that they cancel the $O(1)$ terms in
the Hessian $\Delta$. However, these functions can only change the diagonal
elements of $\Delta$, while the off-diagonal elements  will still have the
$O(1)$ terms.  Note that we have introduced the insertion functions in the
simplest possible way, namely as multiplicative factors for the amplitude of
each link on the boundary spin network. In the most general case a $\mu_l$ can
be a matrix function, see \cite{Mikovic2004}, so that this gives an additional
possibility to change the $O(1)$ behavior. This possibility should be explored,
but the problem is that it is difficult to analyze.

Note that we have assumed that the dominant  contribution to the asymptotics of
the wavefunction comes from a non-degenerate spin foam. The reason was that only
in that case we know how to calculate the asymptotics. Hence there is a
possibility that the dominant contribution comes from a degenerate spin foam and
that this contribution is such that it gives the desired propagator asymptotics.
However, there is a problem with this. Namely, a degenerate spin foam is such
that its every vertex  has at least one small spin. This means that the
corresponding spacetime geometry has the curvature which grately varies at small
scales, which is not consistent with the propagator asymptotics for a flat
spacetime.

The only remaining possibility is to modify the ELPR/FK vertex amplitude
$W(j,\vec{n})$. Note  that the $O(1)$  contribution to $\Delta$ is given by
(\ref{DeltaNula}), and it vanishes if one of the coefficients $N_\pm^{(\alpha)}$
is zero. Consequently, if the modified vertex amplitude $\tilde W(j,\vec{n})$
had the asymptotic behavior
\begin{equation} \label{DobarVerteks}
\tilde W (j,\vec{n}) \approx \frac{e^{i\alpha  S_R^{(v)} (j)}}{V(j)}
\,,\end{equation}
where $V(j)$ is the function from (\ref{jav}), then it is easy to show from
(\ref{OpstiOblikDeltaMatrice}) that 
$$
S = O(1/j_0)\,.
$$
By using (\ref{MuFunkcije}) for the insertion functions, one would then obtain
the correct graviton propagator asymptotics. Note that $\tilde W$ gives a state
sum which for large spins looks like a path integral for Regge discretization of
general relativity, because $\tilde W$ has the asymptotics (\ref{DobarVerteks}).
This explains why $\tilde W$ gives a graviton propagator with a good
asymptotics.  On the other hand, the presence of the complex conjugate term
$e^{-i\alpha S_R^{(v)}}$ in (\ref{jav}) gives an unnatural path integral, so
that it is not a surprise that the corresponding propagator has wrong
asymptotics.

Note that all known spin foam models have the vertex amplitude asymptotics which
is a linear combination of $e^{\pm i\alpha S_R^{(v)}}$ terms, see
\cite{Barret2009eeprl} for the Euclidean ELPR/FK model or \cite{bzc,bs} for the
Barret-Crane model.  Consequently one will obtain $S=O(1)$ for the large-spin
asymptotics of the boundary wavefunction, because the calculation is the same as
the one presented in this paper. It is also instructive to compare our result
with the results of the similar calculation for the Euclidean theory done in
\cite{ConradyFreidel}. This is done in Appendix \ref{AppConradyFreidel}, and
supports our result (\ref{ooa}).

There are two ways to interpret the result (\ref{ooa}). One way is to say that
the choice (\ref{TalasnaFunkcija}) for the boundary wavefunction is not the most
general one, and there may exist another solution which could give the correct
asymptotics. This is of course a possibility, and it is an open problem for
future research. However, such an interpretation of our result essentially
brings us back to the original problem of finding a wavefunction which satisfies
the Hamiltonian constraint and has the asymptotics (\ref{RovelliAnsatz}).
However, it is difficult to see what would be an alternative construction to the
one we used.

The other possibility is to use the same construction for the wavefunction and
to modify the ELPR/FK vertex amplitude, so that the result (\ref{ooa}) is
circumvented.  As discussed above, the way to achieve this is to construct a new
vertex amplitude which would have the asymptotics of the type
(\ref{DobarVerteks}). For example, if $N_+\ne N_-$ then one can define the new
vertex amplitude $\tilde{W}(j,\vec{n})$ as
\begin{equation} \label{ModVertexPrvi}
\tilde W = \frac{ N_+ W - N_- W^*}{N_+^2 - N_-^2} \,,
\end{equation}
where $W^*$ is the complex-conjugate of the ELPR/FK vertex amplitude $W$. The
new amplitude will have the asymptotics (\ref{DobarVerteks}). A more general
redefinition, valid for $N_+ = N_-$ case, is given by
\begin{equation} \label{ModVertexDrugi}
\tilde W = \frac{1}{2N_+} \left( W + \sqrt{W^2 - \frac{4 N_+ N_-}{V^2}} \right)
\,.
\end{equation}
This expression also gives the asymptotics (\ref{DobarVerteks}). Hence the spin
foam model defined by the new amplitude $\tilde W$ will give the correct
propagator asymptotics and it will represent a good candidate for a spin foam
model whose classical limit is general relativity. 

Note that the correct asymptotics of the graviton propagator does not guarantee
that the classical limit of a spin foam model is general relativity. Namely, the
graviton propagator for a boundary state is defined as a 2-point correlation
function. However, in order to determine the corresponding semiclassical
equations of motion one needs the effective action, which is the generating
functional for all $n$-point correlation functions. Knowing just the 2-point
correlation function is not sufficient, so that one needs to compute the
effective action and to show that its classical limit is the Einstein-Hilbert
action.

\begin{acknowledgments}
The authors would like to thank Laurent Freidel for discussions. AM was
partially supported by the FCT grants {\tt PTDC/MAT/69635/2006} and {\tt
PTDC/MAT/099880/2008}. MV was supported by the FCT grant {\tt
SFRH/BPD/46376/2008} and partially by the FCT grant {\tt PTDC/MAT/099880/2008}.
\end{acknowledgments}

\appendix

\section{\label{AppReggeDejstvo}The Regge action for a $4$-simplex}

The Lorentzian Regge action for a $4$-simplex dual to vertex $v$ is given as
\begin{equation} \label{ReggeDejstvo}
S_R^{(v)}(k) = \sum_{f\in v } k_f \Theta_f^{(v)} (k)
\end{equation}
Here $k_f$ are $10$ spins labeling the faces, while each $\Theta_f^{(v)}(k)$ is
the exterior dihedral angle between two tetrahedra of the simplex dual to $v$
which share the triangle dual to $f$.

If all spins $k_f$ are uniformly scaled as $k_f = j_0 \tilde{k}_f$, in the limit
$j_0\to \infty$ the Regge action scales as
\begin{equation} \label{SkaliranjeReggeDejstva}
S_R^{(v)}(k) = O(j_0),
\end{equation}
since $k_f = O(j_0)$ and $\Theta_f^{(v)}(k) = O(1)$.

Also, if we take the derivative of the Regge action with respect to some spin
$k_a$, we obtain
$$
\frac{\del S_R^{(v)} }{\del k_a} = \sum_{f\in v} \delta_{af} \Theta_f^{(v)} + 
\sum_{f\in v} k_f \frac{\del \Theta_f^{(v)} }{\del k_a} \, .
$$
The first sum reduces to $\Theta_a^{(v)}$ if $a\in v$, and is zero otherwise.
The second sum is identically zero due to the Schl\"afli identity, so we have
\begin{equation} \label{IzvodReggeDejstva}
\frac{\del S_R^{(v)} }{\del k_a} = \delta_{a\in v} \Theta_a^{(v)} \, .
\end{equation}
Note that this derivative scales as $\cO(1)$ in the limit $j_0\to\infty$. Also
note that for the nondegenerate $4$-simplex all dihedral angles $\Theta_f^{(v)}$
are different from zero, since the $4$-simplex is always convex. These
properties are essential for the derivation of our results.

\section{\label{AppELPRFKasimptotika}Asymptotics of the ELPR/FK vertex
amplitude}

The asymptotic properties of the ELPR/FK vertex amplitude $W_v$ were
investigated in depth in
\cite{Barret2009ooguri,Barret2009eeprl,Barret2009leprl}, and neatly summarized
in \cite{Barrett2010summarized}.

A single vertex amplitude $W_v$ is a function of $10$ spins $k_f$ and $20$
normals $\vn_{ef}$. Some of the spins may be scaled as $k_f = j_0 \tilde{k}_f$,
while others do not scale. In the limit $j_0\to\infty$, the asymptotic behavior
of $W_v$ can be split into several cases, based on the possible choices of these
variables. These are
\begin{enumerate}
\item \textit{The nondegenerate case} 

In this case we assume all $10$ spins scale with $j_0$, and the boundary of the
corresponding $4$-simplex has the Regge-like geometry. In the case of the
Lorentzian version of the theory, the vertex amplitude has the asymptotic
formula
\begin{equation}
\label{AsimptotikaVerteksa}
W (j_0 k, n) \approx \frac{1}{j_0^{12}} \left[ N_{+}^{(\alpha)} e^{i \alpha j_0
S_R^{(v)}(k)} +
N_{-}^{(\alpha)} e^{-i \alpha j_0  S_R^{(v)}(k)} \right]\,.
\end{equation}
Here $\alpha=1$ for the 4-simplex with a Euclidean geometry on the boundary,
while $\alpha=\gamma$ in the Lorentzian boundary case, where $\gamma$ is the
Immirzi parameter. The constants $N_{\pm}^{(\alpha)}$ are different from zero
and $S_R^{(v)}(k)$ is the Euclidean/Lorentzian Regge action
(\ref{ReggeDejstvo}).

\item \textit{The degenerate cases of zero $4$-volume} 

These are the cases when all $10$ spins scale with $j_0$, but the boundary of
the $4$-simplex does not have Regge-like geometry. The vertex asymptotics was
analyzed in \cite{Barret2009leprl} where it was determined that it has the form
$$
W_v \approx \frac{N(k)}{j_0^{12}}
$$
if the boundary is a $3D$ vector geometry, while
$$
W_v = o(j_0^{-K}), \qquad \forall K \geq 0,
$$
in all other situations with zero $4$-volume. All these cases contribute with
zero measure in in the integral (\ref{EksponencijalniOblikTF}) and can be
ignored.

\item \textit{The degenerate cases of non-zero $4$-volume} 

These are the cases when only some of the $10$ spins scale with $j_0$, while
others are kept fixed. These situations have not been analyzed so far, and the
vertex asymptotics in these cases is still unknown. Note that such
configurations contribute with non-zero measure in the integral
(\ref{EksponencijalniOblikTF}), and thus cannot be ignored.
\end{enumerate}

It is important to emphasize that explicit dependence of the asymptotic formula
on normals $\vn_{ef}$ is lost in (\ref{AsimptotikaVerteksa}), and the asymptotic
expression on the right-hand side of (\ref{AsimptotikaVerteksa}) depends only on
$10$ spins $k_f$. Namely, the assumption of Regge-like geometry of the
$4$-simplex implies that its triangle areas $k_f$ and its normals $\vn_{ef}$ are
fully determined by its $10$ edge lengths $l_i$, which also induce the
Lorentzian/Euclidean signature of the metric in the $4$-simplex. However, given
that the number of triangles in a $4$-simplex is equal to its number of edges,
the functions $k_f(l_i)$ can in a generic situation be inverted, and edge
lengths regarded as functions of the triangle areas. This is possible always
except in some particular cases where the Jacobian of the transformation is
singular. Nevertheless, these singular cases contribute with zero measure in the
integral (\ref{EksponencijalniOblikTF}) and can thus be ignored. Given the
inverted functions $l_i(k_f)$, one can also express the normals $\vn_{ef}(l_i)$
as functions of $k_f$, which therefore remain the only independent variables in
(\ref{AsimptotikaVerteksa}).

Note that the asymptotics (\ref{AsimptotikaVerteksa}) can be rewritten as
\begin{equation} 
W (j, n) \approx \frac{1}{V(j)} \left[ N_{+}^{(\alpha)} e^{i \alpha 
S_R^{(v)}(j)} +
N_{-}^{(\alpha)} e^{-i \alpha  S_R^{(v)}(j)} \right]\,,\label{jav}
\end{equation}
for $j\to\infty$ where $V(j) = O(j^{12})$.

\section{\label{AppendixTeoremaSmatrice}Proof of Theorem 1}

Here we give a proof of Theorem 1 used in the main text. Let us repeat the
statement of the theorem, for completeness.

\medskip

\textbf{Theorem 1.} \textit{If the matrix $S$ is nonzero, and if the leading
order contribution to $\Delta$ comes from $W_v$ terms, we have}
$$
S = O(\Delta)
$$
\textit{for $j_0 \to\infty$. If the matrix $S$ is zero, the wavefunction
asymptotics is non-Gaussian.}

\medskip

The proof goes as follows. Begin by noting that the Hessian matrix $\Delta$ is
non-diagonal. Namely, looking at its definition (\ref{DefinicijaDeltaMatrice})
and the action (\ref{Dejstvo}), we see that the insertion functions $\mu_l$ and
face amplitudes $d_f$ contribute only to diagonal terms in $\Delta$, since each
of them is a function of a single variable. In contrast to this, the vertex
amplitudes $W_v$ are functions of $10$ or $30$ variables each, according to the
combinatorics of the spin foam $2$-complex and the possible degeneracy of $W_v$.
Therefore, each vertex amplitude will contribute to the diagonal terms of
$\Delta$, and in addition also to some off-diagonal terms on each side of the
main diagonal, in such a way that in every row and column there will be some
nonzero nondiagonal elements present.

We want to discuss the dependence of $\det\Delta$ on $j_0$ in the limit
$j_0\to\infty$. For simplicity, in what follows we shall assume that $\det
\Delta \neq 0$, and we shall discuss the singular case later.

The determinant of $\Delta$ is by definition given as
$$
\det \Delta = \sum_p \sgn (p) \Delta_{1p(1)} \Delta_{2p(2)} \dots \Delta_{Rp(R)}
\,,
$$
where $p$ is the permutation of indices $1\dots R$, and $R$ is the rank (and
simultaneously the dimension) of $\Delta$. In this sum, there will be some terms
which contain diagonal terms of $\Delta$, and terms which do not contain any
diagonal element. The first set of terms will have contributions of $\mu_l$,
$d_f$ and $W_v$, while the second set of terms will be determined solely by
amplitudes $W_v$. Given the assumption that the leading order of $\Delta$ comes
from $W_v$, we have that the determinant of $\Delta$ will scale with $j_0$ as:
$$
\det \Delta = \cO(\Delta^R).
$$
Namely, the scaling of terms with diagonal elements in the determinant cannot be
established without the detailed knowledge of its dependence on $\mu_l$ and
$d_f$ terms. However, the scaling of each off-diagonal component
$\Delta_{kp(k)}$ (where $p(k)\neq k$, $k=1,\dots,R$) will be determined only by
the vertex amplitude $W_v$, and is dominant by assumption. As a consequence, the
terms in $\det \Delta$ which do not contain any diagonal elements are dominant
and scale as $\cO(\Delta^R)$, while the terms which do contain diagonal elements
may scale with smaller power in $j_0$ and can be neglected in the limit
$j_0\to\infty$.

Once the scaling of $\det \Delta$ has been established, we can employ some
well-known results about the Schur complement matrix in order to establish the
scaling of $S$. These results are summarized and proved in the form of Lemma 1
in Appendix \ref{AppendixMatrixLemma}.

Let the Hessian matrix $\Delta$, its submatrix $M$ and its Schur complement
$\tS$ scale as
$$
\Delta = \cO \left( \frac{1}{j_0^d}\right) , \qquad M = \cO \left(
\frac{1}{j_0^d}\right),
\qquad \tS = \cO \left( \frac{1}{j_0^s}\right).
$$
Note that $M$, being the submatrix of $\Delta$, scales with $j_0$ with the same
power $-d$ as $\Delta$. However, this cannot be assumed for the Schur complement
$\tS$ since there might be nontrivial cancellations between the leading terms in
$A$ and $NM^{-1}N^T$ in (\ref{SchurovKomplement}). Consequently, the scaling
power of $\tS$ is $-s$. What we need to prove is that these cancellations do not
happen, and that in fact $s=d$.

Denote the ranks of $M$ and $\tS$ matrices as $r$ and $\rho$, respectively. By
part (b) of the Lemma 1, we have
$$
\det \Delta = \det M \det \tS.
$$
Calculating the scaling order of the left-hand and right-hand sides, and using
the usual properties of determinants, we easily see that
$$
\frac{1}{j_0^{Rd}} = \frac{1}{j_0^{rd}} \frac{1}{j_0^{\rho s}},
$$
which gives
$$
Rd=rd+\rho s.
$$
By the part (a) of the Lemma, we have $R = r + \rho$. Using this to eliminate
both $R$ and $r$, the above equation reduces to
$$
\rho (s-d) = 0 \, .
$$
Finally, by assumption of the theorem, matrix $\tS$ is nonzero, which means that
its rank $\rho$ is nonzero. Therefore we conclude that $s=d$, which actually
means that $\tS = \cO(\Delta)$. As matrix $S$ is a $jj$ submatrix of $\tS$, it
scales in the same way as $\tS$. Consequently,
$$
S = \cO(\Delta)\, ,
$$
which proves the theorem in the case when $\Delta$ is nondegenerate.

If $\Delta$ has zero eigenvalues, the determinant equation above vanishes
identically. However, in this case we can repeat the whole analysis in the same
way, except that we need to use part (c) of the Lemma instead of part (b),
bearing in mind that $\cO(B_4)=1$ (see Remark 3 in Appendix
\ref{AppendixMatrixLemma}). Namely, instead of analyzing the determinants of
$\Delta$, $M$ and $\tS$, we can rotate the basis to represent these three
matrices in the form
$$
\Delta = \left[
\begin{array}{cc}
0 & 0 \\
0 & M_{\Delta} \\
\end{array}
\right], \quad
M = \left[
\begin{array}{cc}
0 & 0 \\
0 & \tM \\
\end{array}
\right], \quad
\tS = \left[
\begin{array}{cc}
0 & 0 \\
0 & M_{\tS} \\
\end{array}
\right] \, ,
$$
and repeat the whole proof using the regular minors $M_{\Delta}$, $\tM$ and
$M_{\tS}$ instead. Note that as a consequence of the part (a) of the Lemma, the
sum of dimensions of the zero-blocks of $M$ and $\tS$ must be equal to the
dimension of the zero-block of $\Delta$. These zero-blocks represent the kernel
of $\Delta$, and as discussed in the main text, appear as a consequence of
continuous symmetries of the action (\ref{Dejstvo}). As was shown in section
\ref{SekV}, they may safely be integrated out, and the Schur complement
(\ref{SchurovKomplement}) constructed from the regular part of $M$, i.e. the
minors $\tM$ and $\tN$.

Again, since $S$ is a submatrix of $\tS$, if it is nonzero it scales with the
same power as $\tS$, so consequently we have
$$
S = \cO(\Delta)\, ,
$$
in the degenerate case as well. This completes the proof of the theorem.

\section{\label{AppendixMatrixLemma}Properties of the Schur complement}

Here we establish some properties of the Schur complement that we have used in
the proof of Theorem 1. These results can be found in \cite{Zhang2005}. However,
one of the results, the statement (c) below, is a new result, generalizing the
statement (b).

\medskip

\textbf{Lemma 1.} \textit{Let $\Delta$ be a symmetric complex matrix of type
$n\times n$ and let $R$ be its rank. Let us split $\Delta$ into blocks as
$$
\Delta = \left[
\begin{array}{cc}
A & N \\
N^T & M \\
\end{array}
\right],
$$
where $A$ is a $J\times J$ matrix, $N$ is a $J\times r$ matrix, $M$ is a
$r\times r$ matrix  and $n = J + r$. We will also assume that $M$ is invertible,
and that real and imaginary parts of $\Delta$ commute.}

\textit{Let us construct the Schur complement $\tS$ (see}
\cite{Zhang2005}\textit{), which is a $J\times J$ matrix
$$
\tS = A - N M^{-1} N^T.
$$
Denote the rank of $\tS$ as $\rho$. Then
\begin{itemize}
\item[(a)] $R = r + \rho$ (Guttman rank additivity);
\item[(b)] $\det \Delta = \det \tS \det M$ (Schur determinant formula);
\item[(c)] if $0< \rho < J$, then
\begin{equation} \label{IdentitetZaDeterminante}
\det M_{\Delta} (\det B_4)^2 = \det M \det M_{\tS}.
\end{equation}
\end{itemize}
}

\textit{Here $M_{\Delta}$ and $M_{\tS}$ are invertible $R\times R$ and
$\rho\times\rho$ matrices, respectively. They are obtained by using orthogonal
transformations which put $\Delta$ and $\tS$ into a block-diagonal form
$$
\Delta = \left[
\begin{array}{cc}
0 & 0 \\
0 & M_{\Delta} \\
\end{array}
\right], \qquad
\tS = \left[
\begin{array}{cc}
0 & 0 \\
0 & M_{\tS} \\
\end{array}
\right],
$$
The $B_4$ matrix will be explicitly constructed in the proof below.}

\medskip

\textit{Proof}. We start from the Aitken block diagonalization formula
\cite{Zhang2005} and from now on we use $I$ to denote a unit matrix of any size
appropriate for its position in an equation:
 \begin{equation} \label{AitkenFormula}
\left[
\begin{array}{cc}
I & -NM^{-1} \\
0 & I \\
\end{array}
\right] \!
\left[
\begin{array}{cc}
A & N \\
N^T & M \\
\end{array}
\right] \!
\left[
\begin{array}{cc}
I & 0 \\
-M^{-1}N^T & I \\
\end{array}
\right] \!
=
\! \left[
\begin{array}{cc}
\tS & 0 \\
0 & M \\
\end{array}
\right] .
\end{equation}

This equation can be verified by a direct multiplication of the left-hand side.
Denoting the first matrix on the left as $C$, we can rewrite this identity in a
compact form $C\Delta C^T = \tS \oplus M$. The rank of the right-hand side is
the sum of ranks of $\tS$ and $M$, which amounts to $\rho+r$. Since the rank of
$C$ is equal to its dimension $n$, the total rank of the product on the
left-hand side is equal to the rank of $\Delta$, so we easily obtain
$$
R = r + \rho\,,
$$
which completes the proof of part (a).

Next, we take the determinant of (\ref{AitkenFormula}). Since $C$ is
block-triangular, its determinant is a product of determinants of blocks on the
diagonal, so we obtain $\det C =1$. The left-hand side is thus the product of
determinants, $\det C \det \Delta \det C^T$, and it is equal to $\det\Delta$
because $\det C^T=\det C =1$. On the right-hand side we have a block-diagonal
matrix, so that its determinant is equal to $\det\tS \det M$. Hence,
$$
\det \Delta = \det\tS \det M \,,
$$
which completes the proof of part (b).

In order to prove (c), let $O$ be a $J\times J$ orthogonal matrix which
transforms $\tS$ into a block-reduced form,
$$
O\tS O^T = 0\oplus M_{\tS}\,.
$$
Since $\rho\neq 0$, matrix $\tS$ has exactly $\rho$ nonzero eigenvalues, which
constitute $M_{\tS}$. Given that the eigenvalues of $M_{\tS}$ are nonzero, it is
invertible. The zero-block is of type $\nu\times\nu$, where $\nu = J-\rho$ is
the dimension of the null-space of $\tS$. By using $O$ one can construct an
orthogonal $n\times n$ matrix $P = O\oplus I$ such that
\begin{equation} \label{MatricaP}
P \left( \tS \oplus M \right) P^T = 0\oplus M_{\tS} \oplus M.
\end{equation}

By using an analogous argument one can always construct an orthogonal $n\times
n$ matrix $Q^T$ such that
$$
Q^T\Delta Q = 0\oplus M_{\Delta},
$$
which can be solved for $\Delta$:
\begin{equation} \label{MatricaQ}
\Delta = Q \left( 0\oplus M_{\Delta} \right) Q^T \,.
\end{equation}
The zero block comes from the null-space of $\Delta$. It is of the size $n-R$,
which is also equal to $\nu$, since $n=J+r$ and $R = r+\rho$ according to the
part (a).

Consider (\ref{AitkenFormula}), and multiply it by $P$ from the left and by
$P^T$ from the right, and use (\ref{MatricaP}) and (\ref{MatricaQ}) to rewrite
it in the form
\begin{equation} \label{VezaMinoraZaDeltaIsTilda}
PCQ \left( 0\oplus M_{\Delta} \right) Q^T C^T P^T = 0\oplus M_{\tS} \oplus M \,.
\end{equation}

Let us introduce the matrix $B\equiv PCQ$ and write it in the block form as
$$
B = \left[
\begin{array}{cc}
B_1 & B_2 \\
B_3 & B_4 \\
\end{array}
\right],
$$
where the blocks $B_1$, $B_2$, $B_3$ and $B_4$ are $\nu\times\nu$, $\nu\times
R$, $R\times\nu$ and $R\times R$ matrices, respectively. Substituting this into
the left-hand side of (\ref{VezaMinoraZaDeltaIsTilda}) yields
$$
PCQ \left( 0\oplus M_{\Delta} \right) Q^T C^T P^T \equiv
B \left[
\begin{array}{cc}
0 & 0 \\
0 & M_{\Delta} \\
\end{array}
\right] B^T =
$$
\begin{equation} \label{LevaStranaMatricneJednacine}
= \left[
\begin{array}{cc}
B_2 M_{\Delta} B_2^T & B_2 M_{\Delta} B_4^T \\
B_4 M_{\Delta} B_2^T & B_4 M_{\Delta} B_4^T \\
\end{array}
\right].
\end{equation}

By comparing (\ref{LevaStranaMatricneJednacine}) to the right-hand side of
(\ref{VezaMinoraZaDeltaIsTilda}), we obtain
\begin{equation} \label{IzjednacenaLevaIdesnaStrana}
\left[
\begin{array}{cc}
B_2 M_{\Delta} B_2^T & B_2 M_{\Delta} B_4^T \\
B_4 M_{\Delta} B_2^T & B_4 M_{\Delta} B_4^T \\
\end{array}
\right]
=
\left[
\begin{array}{ccc}
0 & 0 & 0 \\
0 & M_{\tS} & 0 \\
0 & 0 & M \\
\end{array}
\right]\,.
\end{equation}

Note that the zero-block of (\ref{IzjednacenaLevaIdesnaStrana}) is a
$\nu\times\nu$ matrix, which is also the $B_2 M_{\Delta} B_2^T$ block. We then
read off the following equations
\begin{equation} \label{VezaMovaIbeCetiri}
B_4 M_{\Delta} B_4^T = M_{\tS} \oplus M\,,
\end{equation}
\begin{equation} \label{JednacinaZaBeDva}
B_2 M_{\Delta} B_4^T = 0\,,
\end{equation}
\begin{equation} \label{DrugaJednacinaZaBeDva}
B_2 M_{\Delta} B_2^T = 0\,.
\end{equation}

By taking the determinant of (\ref{VezaMovaIbeCetiri}), we finally obtain
$$
\det M_{\Delta} (\det B_4)^2 = \det M \det M_{\tS} \,.
$$
This establishes (\ref{IdentitetZaDeterminante}) and completes the proof of part
(c).

Given that $M$, $M_{\tS}$ and $M_{\Delta}$ are all invertible, we have $\det B_4
\neq 0$ which means that $B_4$ is also invertible. By multiplying
(\ref{JednacinaZaBeDva}) by $(B_4^T)^{-1}M_{\Delta}^{-1}$ from the right, we
obtain
$$
B_2 = 0.
$$
The equation (\ref{DrugaJednacinaZaBeDva}) now vanishes and does not provide any
additional constraint. Therefore, the matrix $B$ has the following form
\begin{equation} \label{OblikMatriceB}
B \equiv PCQ = \left[
\begin{array}{cc}
B_1 & 0 \\
B_3 & B_4 \\
\end{array}
\right] \,.
\end{equation}
\textit{End of proof}.

\medskip

\textbf{Remark 1.} The $\Delta$ matrix from the main text has the form
$$
\Delta = \left[
\begin{array}{ccc}
A & N & 0 \\
N^T & M & 0 \\
0 & 0 & 0 \\
\end{array}
\right],
$$
which differs from the one in Lemma 1 by an additional zero-block. However,
these additional zeroes are integrated out before the lemma is applied, and
hence they do not affect any statements of lemma.

\medskip

\textbf{Remark 2.} The result (c) is a generalization of the result (b) to the
case when $\Delta$ is a singular matrix. While the part (b) is in fact valid for
singular matrices, it merely states that $0=0$ and provides no information about
nonsingular principal minors of $\Delta$. The result (c) is more fine-grained,
and provides precisely this nontrivial information about $\Delta$.

It was assumed in the part (c) that $0<\rho < J$. If $\rho = J$ then $\Delta$ is
a regular matrix, and hence the result (b) can be used. If $\rho = 0$, then $\tS
= 0$, $\nu = J$, and instead of (\ref{VezaMovaIbeCetiri}) we obtain
$$
B_4 M_{\Delta} B_4^T = M \,,
$$
and consequently
$$
\det M_{\Delta} (\det B_4)^2 = \det M\, .
$$

In this case we can set $P=I$ and obtain
$$
B \equiv CQ = \left[
\begin{array}{cc}
B_1 & 0 \\
B_3 & B_4 \\
\end{array}
\right]
$$
for the matrix $B$.

\medskip

\textbf{Remark 3.} In Appendix \ref{AppendixTeoremaSmatrice} we use the results
(b) and (c) to determine the leading $j_0$-order of the Schur complement $\tS$,
knowing $\cO(\Delta)$. However, it is necessary to show that $B_4$ is of order
$\cO(1)$. In order to do this, note that
$$
\det B = \det P \det C \det Q = \pm 1 \,,
$$
since $P$ and $Q$ are orthogonal matrices. On the other hand, from
(\ref{OblikMatriceB}) we know that $\det B = \det B_1 \det B_4$, so that we have
\begin{equation} \label{VezaDeterminantiBjedanIBcetiri}
\det B_1 \det B_4 = \pm 1 \,.
\end{equation}

Let us now assume that the blocks $B_1$ and $B_4$ are of order $k$ and $m$ in
$1/j_0$, respectively
$$
B_1 = \frac{D}{j_0^k} + \cO\left( \frac{1}{j_0^{k+1}} \right), \qquad
B_4 = \frac{E}{j_0^m} + \cO\left( \frac{1}{j_0^{m+1}} \right),
$$
$$
k,m\geq 0, \qquad D,E \sim \cO(1) \,.
$$
The numbers $k$ and $m$ cannot be negative since the whole $B$ matrix must be of
order $\cO(1)$. Namely, the matrices $P$ and $Q$ are orthogonal, and
consequently all their elements are bounded above by $1$. Thus $P$ and $Q$ are
$\cO(1)$. The matrix $C$ is also $\cO(1)$, since $\Delta$ and consequently $M$,
$N$, $M^{-1}$ are all of the same order. Therefore, $B = PCQ \sim \cO(1)$.

Since $B_1$ is a $\nu\times\nu$ matrix and $B_4$ is a $R\times R$ matrix, we
have
\begin{subequations} \label{DeterminanteBjedanIBcetiri}
\begin{equation} \label{DeterminanteBjedanIBcetiriA}
\det B_1 = \frac{1}{j_0^{k\nu}} \det D + \cO\left( \frac{1}{j_0^{k+1}}
\right)\,,
\end{equation}
\begin{equation} \label{DeterminanteBjedanIBcetiriB}
\det B_4 = \frac{1}{j_0^{mR}} \det E + \cO\left( \frac{1}{j_0^{m+1}}
\right)\,.
\end{equation}
\end{subequations}

Substituting (\ref{DeterminanteBjedanIBcetiri}) back into
(\ref{VezaDeterminantiBjedanIBcetiri}) we obtain the consistency equation
$$
k\nu + mR = 0 \,.
$$
Since both $\nu,R > 0$ while $k,m\geq 0$, the only solution of this equation is
$k=m=0$. Therefore
$$
\det B_4 \sim B_4 \sim \cO(1) \,.
$$

In the case when $\nu=0$ the $\Delta$ matrix is regular and instead of the part
(c) we use the part (b) of Lemma 1. However, the part (b) does not involve $\det
B_4$, so that we need the above result only for $\nu>0$.

\section{\label{AppIzvodiVerteksa}Vertex amplitude contribution to the Hessian
matrix}

Here we give the explicit formulae for the terms on the right-hand side of
(\ref{DeltaV}). These terms are calculated by directly substituting the vertex
asymptotics (\ref{AsimptotikaVerteksa}) into equation
(\ref{OpstiOblikDeltaMatrice}) and differentiating. It is important to note that
in the expression (\ref{AsimptotikaVerteksa}) the $x$-dependence is in the
coefficients $N_+$ and $N_-$, as well as in the Regge action $S_R$. However, the
scaling of $N_{\pm}$ is different than that of $S_R$. The former scale as $O(1)$
while the latter scales as $O(j_0)$ in the limit $j_0\to\infty$.

The $O(1)$ term in (\ref{DeltaV}) has already been quoted in the text in
equation (\ref{DeltaNula}), and we repeat it here for completeness:
\begin{widetext}
$$
\Delta_{vab}^{(0)} =
 \left[ \left(
\frac{N_+^{(\alpha)} e^{i\alpha S_R^{(v)}} - N_-^{(\alpha)} e^{-i\alpha
S_R^{(v)}}}
{N_+^{(\alpha)} e^{i\alpha S_R^{(v)}} + N_-^{(\alpha)} e^{-i\alpha S_R^{(v)}}}
\right)^2 -1 \right] \alpha^2 \delta_{a,b\in v} \Theta_a^{(v)} \Theta_b^{(v)} \,
.
$$
The $O(j_0^{-1})$ term is given as:
$$
\Delta_{vab}^{(1)} = \frac{N_+^{(\alpha)} e^{i\alpha S_R^{(v)}} -
N_-^{(\alpha)} e^{-i\alpha S_R^{(v)}}}{N_+^{(\alpha)} e^{i\alpha S_R^{(v)}} +
N_-^{(\alpha)} e^{-i\alpha S_R^{(v)}}}
i\alpha \delta_{a,b\in v} \del_a \Theta_b^{(v)} +
$$
$$
+
2\frac{N_-^{(\alpha)} \del_a N_+^{(\alpha)} - N_+^{(\alpha)} \del_a
N_-^{(\alpha)}
}{\left(N_+^{(\alpha)} e^{i\alpha S_R^{(v)}} +
N_-^{(\alpha)} e^{-i\alpha S_R^{(v)}} \right)^2 } i\alpha 
\delta_{a,b\in v} \Theta_b^{(v)}
+
2\frac{N_-^{(\alpha)} \del_b N_+^{(\alpha)} - N_+^{(\alpha)} \del_b
N_-^{(\alpha)}
}{\left(N_+^{(\alpha)} e^{i\alpha S_R^{(v)}} +
N_-^{(\alpha)} e^{-i\alpha S_R^{(v)}} \right)^2 } i\alpha 
\delta_{a,b\in v} \Theta_a^{(v)} \, .
$$
The $O(j_0^{-2})$ term is given as:
$$
\Delta_{vab}^{(2)} =
\frac{ e^{i\alpha S_R^{(v)}} \del_a\del_b N_+^{(\alpha)} +
 e^{-i\alpha S_R^{(v)}} \del_a \del_b N_-^{(\alpha)} }{N_+^{(\alpha)} e^{i\alpha
S_R^{(v)}} +
N_-^{(\alpha)} e^{-i\alpha S_R^{(v)}}}
\delta_{a,b\in v} -
$$
$$
-\frac{ \left( e^{i\alpha S_R^{(v)}} \del_a N_+^{(\alpha)} +
 e^{-i\alpha S_R^{(v)}} \del_a N_-^{(\alpha)} \right)
\left( e^{i\alpha S_R^{(v)}} \del_b N_+^{(\alpha)} +
 e^{-i\alpha S_R^{(v)}} \del_b N_-^{(\alpha)} \right)
}{\left(N_+^{(\alpha)} e^{i\alpha S_R^{(v)}} +
N_-^{(\alpha)} e^{-i\alpha S_R^{(v)}} \right)^2 }
\delta_{a,b\in v} \, .
$$
\end{widetext}

\section{\label{AppConradyFreidel}Comparison with the Euclidean results}

It is instructive to compare our results with the asymptotic analysis of the
Euclidean ELPR/FK state-sum kernel given in \cite{ConradyFreidel}. There are
several differences in the setting between the approach of \cite{ConradyFreidel}
and the one taken in this paper.  Since they do not consider the boundary
wavefunction, they do not have the insertion functions. Next, they integrate the
state sum over all variables except the spins, and obtain
$$
Z_{\Delta} = \sum_{j_f} \prod_f d_{j_f^{\gamma+}} d_{j_f^{\gamma-}}
W_{\Delta}(j_f)\,,
$$
where $W_{\Delta}(j_f)$ is the stat-sum kernel, $\Delta$ is the triangulation
dual to $\sigma$ and the face amplitude is a product of two $d_f$ terms.

One of the main results of \cite{ConradyFreidel} is the asymptotic large-spin
expression for the nondegenerate part of this kernel
$$
W_{\Delta}^{\ndg}(N j_f) = \frac{c_{\Delta}(j_f)}{N^{\frac{r_{\Delta}}{2}}} \cos
(N S_R) \, ,
\qquad N \to \infty \, ,
$$
where $N$ is the large parameter, $c_{\Delta}$ is the function of spins, but not
of $N$ and $S_R$ is
$$
S_R = (\gamma^+ + \gamma^-) \sum_f j_f \Theta_f\, , \qquad \Theta_f = \sum_{e\in
f} \theta_{ef}\,,
$$
see equations (87) and (101) in \cite{ConradyFreidel}. Note that $\Theta_f$ is
the sum of all dihedral angles around a face $f$. The action $S_R$ is
constructed for the triangulation $\Delta$, with non-scaled spins, and the large
parameter $N$ is written explicitly in front of it in $W_{\Delta}$. In our
notation, the above expression can be rewritten as
$$
W_{\Delta}^{\ndg}(j) = \cA(j) \cos (S_R) \, ,
\qquad j \to \infty \, ,
$$
where now $S_R$ is constructed with the scaled spins $j$, while the amplitude is
denoted simply by $\cA(j)$.

Despite all the differences in the two setups, there is a rather simple generic
relation between the kernel $W_{\Delta}(j)$ and our boundary wavefunction
(\ref{TalasnaFunkcija}). The quantity that corresponds to
(\ref{TalasnaFunkcija}) can be constructed from the kernel $W_{\Delta}(j)$ in
the following way. First we choose the triangulation $\Delta$ so that it has a
boundary. The dual of $\Delta$ will be the 2-complex $\sigma$, while the dual of
the boundary will be a 1-complex $\gamma = \del\sigma$. Next, we split the face
labels $j_f$ into the boundary and internal labels, and denote them $j_f$ and
$k_f$, respectively. Then the wavefunction is given as
$$
\Psi_{\gamma}^{\ndg}(j) = \sum_{k_f} \prod_f (d_{+} d_{-})
W_{\Delta}^{\ndg}(j,k)\, .
$$
Here we have not introduced the insertion functions $\mu_l$ on the boundary, and
the face amplitude is quadratic in spins.

In the limit where both internal and boundary spins are large, one can
approximate the sum with an integral and write the wavefunction as
\begin{equation} \label{StateSumaUConFridRadu}
\Psi_{\gamma}^{\ndg}(j) = \int dk \prod_f (d_{+} d_{-}) \cA(j,k) \cos
(S_R(j,k))\, .
\end{equation}
In order to evaluate the integral over $k$ in the large-spin limit, we would
like to approximate the cosine with a Gaussian in the neighborhood of each of
its stationary points, and employ the stationary-point method. This technique
was used for the asymptotic analysis of the wavefunction for the Euclidean
canonical LQG \cite{Mikovic2010}, which is given by a similar state sum as
(\ref{StateSumaUConFridRadu}). The cosine has infinitely many stationary points,
but we can assume that the amplitude $\cA(j,k)$ is peaked around only one of
them (otherwise the asymptotics will never be a Gaussian function; also one
could introduce the insertion functions which would single out one stationary
point). Denote this stationary point as $(j^*,k^*)$. In the neighborhood of this
point, the cosine can be approximated with a Gaussian via the formula
$$
\cos S_R(x) = e^{\log\cos S_R(x)} \approx e^{-\frac{1}{2}(x-x^*)^T \Delta
(x-x^*)} \,,
\qquad x\to x^* \, .
$$
where $x=(j,k)$ and $\Delta_{ab} \equiv \frac{\del^2}{\del x_a \del x_b}
\log\cos S_R(x)\Big|_{x^*}$. Using the fact that $x^*$ is the stationary point
of the cosine and the fact that the variation of $S_R$ w.r.t. the angles
$\Theta_f$ is identically zero, an explicit evaluation of $\Delta_{ab}$ gives
$$
\Delta_{ab} = -\Theta_a\Theta_b = O(1) \, .
$$

At this point one can split the matrix $\Delta$ into $jj$, $jk$ and $kk$ blocks,
like in Eq. (\ref{BlokoviDeltaMatrice}) and perform the integration over
$k$-spins. The result will be a Gaussian over the remaining $j$-spins,
$$
\Psi_{\gamma}^{\ndg}(j) = \cA(j^*) e^{-\frac{1}{2}(j-j^*)^T S (j-j^*)} \,,
$$
where $S$ is the Schur complement of $\Delta$. Since the order of $\Delta$ is
$O(1)$, so will be the order of the matrix $S$, confirming our general result
(\ref{ooa}).


\begin{thebibliography}{99}

\bibitem{lqg}
C. Rovelli,
Quantum Gravity,
{\it Cambridge University Press}, Cambridge (2004).

\bibitem{Mikovic2004}
A. Mikovi\'c,
{\it Class. Quant. Grav.} {\bf 21} 3909 (2004),
{\tt arXiv:gr-qc/0404021}.
Errata: {\it Class. Quant. Grav.} {\bf 23} 5459 (2006),
{\tt arXiv:gr-qc/0606081}.

\bibitem{Rovelli2006}
C. Rovelli,
{\it Phys. Rev. Lett.} {\bf 97} 151301 (2006),
{\tt arXiv:gr-qc/0508124}.

\bibitem{Bianchi2006}
E. Bianchi, L. Modesto, C. Rovelli and S, Speziale,
{\it Class. Quant. Grav.} {\bf 23} 6989 (2006),
{\tt arXiv:gr-qc/0604044}.

\bibitem{Alesci2008}
E. Alesci and C. Rovelli,
{\it Phys. Rev. D}  {\bf 77} 044024 (2008),
{\tt arXiv:0711.1284}.

\bibitem{Mikovic2008}
A. Mikovi\'c,
{\it Fortschr. Phys.} {\bf 56} 475 (2008),
{\tt arXiv:0706.0466}.

\bibitem{Bianchi2009}
E. Bianchi, E. Magliaro and C. Perini,
{\it Nucl. Phys.} {\bf B822} 245 (2009),
{\tt arXiv:0905.4082}.

\bibitem{Freidel}
B. Dittrich, L. Freidel and S. Speziale,
{\it Phys. Rev. D} {\bf 76} 104020 (2007),
{\tt arXiv:0707.4513}.

\bibitem{Mikovic2010}
A. Mikovi\'c and M. Vojinovi\'c,
to appear in {\it Adv. Theor. Math. Phys.},
{\tt arXiv:1005.1866}.

\bibitem{elpr}
J. Engle, E. Livine, R. Pereira and C. Rovelli,
{\it Nucl. Phys.} {\bf B799} 136 (2008),
{\tt arXiv:0711.0146}.

\bibitem{fk}
L. Freidel and K. Krasnov,
{\it Class. Quant. Grav.} {\bf 25} 125018 (2008),
{\tt arXiv:0708.1595}.

\bibitem{HartleHawking}
J. B. Hartle and S. W. Hawking,
{\it Phys. Rev. D} {\bf 28}, 2960 (1983).

\bibitem{cohst}
E. R. Livine and S. Speziale,
{\it Phys. Rev. D} {\bf 76} 084028 (2007),
{\tt arXiv:0705.0674}.

\bibitem{Rovelli2010}
C. Rovelli,
{\tt arXiv:1004.1780}.

\bibitem{Bianchi2010}
E. Bianchi, D. Regoli and C. Rovelli,
{\tt arXiv:1005.0764}.

\bibitem{Barret2009eeprl}
J. Barrett, R. Dowdall, W. Fairbairn, H. Gomes and F. Hellmann,
{\it J. Math. Phys.} {\bf 50} 112504 (2009),
{\tt arXiv:0902.1170}.

\bibitem{Hormander1983}
L. H\"ormander,
The Analysis of Linear Partial Diﬀerential Operators I,
{\it Springer-Verlag}, Berlin (1983).

\bibitem{ConradyFreidel}
F. Conrady and L. Freidel,
{\it Phys. Rev. D} {\bf 78} 104023 (2008),
{\tt arXiv:0809.2280}.

\bibitem{MagliaroPerini}
E. Magliaro and C. Perini,
{\tt arXiv:1105.0216}

\bibitem{Barret2009ooguri}
J. Barrett, W. Fairbairn and F. Hellmann,
{\tt arXiv:0912.4907}.

\bibitem{Barret2009leprl}
J. Barrett, R. Dowdall, W. Fairbairn, F. Hellmann and R. Pereira,
{\tt arXiv:0907.2440}.

\bibitem{Zhang2005}
F. Zhang,
The Schur Complement and its Applications,
{\it Springer}, New York (2005).

\bibitem{bzc}
J. Baez, J. Christensen and G. Egan,
{\it Class. Quant. Grav.} {\bf 19} 6489 (2002),
{\tt arXiv:gr-qc/0208010}.
 
\bibitem{bs}
J. Barrett and C. Steele,
{\it Class. Quant. Grav.} {\bf 20} 1341 (2003),
{\tt arXiv:gr-qc/0209023}.

\bibitem{Barrett2010summarized}
J. Barrett, R. Dowdall, W. Fairbairn, H. Gomes, F. Hellmann and R. Pereira,
{\tt arXiv:1003.1886}.






\end{thebibliography}
\end{document}